# Mesoscopic quantum coherence in an optical lattice

D. L. Haycock (1), P. M. Alsing (2), I. H. Deutsch (2), J. Grondalski (2)
and P. S. Jessen (1)

*(1) Optical Sciences Center, University of Arizona, Tucson, Arizona 85721*
*(2) Department of Physics and Astronomy, University of New Mexico, Albuquerque, New Mexico 87131.*

We observe the quantum coherent dynamics of atomic spinor wavepackets in the double well potentials of a far-off-resonance optical lattice. With appropriate initial conditions the system Rabi oscillates between the left and right localized states of the ground doublet, and at certain times the wavepacket corresponds to a coherent superposition of these mesoscopically distinguishable quantum states. The atom/optical double well potential is a flexible and powerful system for further study of mesoscopic quantum coherence, quantum control and the quantum/classical transition.

Quantum coherence between localized but separated states of a particle in a double-well potential has long served as a paradigm for nonclassical dynamics. Of particular interest is the possibility to create and manipulate coherent superpositions of meso- or macroscopically distinct quantum states, and to study the role played by decoherence in the emergence of classical dynamics [1]. Extending the limits of coherent control of large quantum systems is of great fundamental interest, and lies at the heart of the quest for quantum computation [2] and related areas of quantum information science. Macroscopic quantum tunneling [3], i. e. the *incoherent* decay of a metastable quantum state along some macroscopic system coordinate, is known to occur for the phase difference of the superconducting order parameter across Josephson junctions [4], and has been observed also in the relaxation of the cooperative magnetization vector in magnetic grains [5] and in the relaxation of spin-domains in atomic Bose-Einstein Condensates (BECs) [6]. Such phenomena do not provide evidence for the existence of superpositions of macroscopically distinguishable quantum states, for which one must undertake the much harder challenge of demonstrating *coherent* dynamics on the macroscopic scale. So far, coherent quantum dynamics has been achieved only on a mesoscopic scale, notably in ion traps [7] and cavity QED [8], though proposals have been put forward to search for macroscopic quantum coherence with SQUIDs [9] and BECs [10].

In this article we report the observation of coherent dynamics in a new mesoscopic quantum system consisting of Cesium atoms in wavelength-sized optical double-well potentials in a far-off-resonance optical lattice. The atomic wavepackets undergo clear Rabi oscillations between two localized states separated by ~150 nm, at frequencies that show excellent quantitative agreement with calculations using no free parameters. Our experiment gives practical insight into the decoherence and dephasing of delocalized atomic wavepackets in very deep optical lattices, and provides guidance for recent proposals to implement qubits and quantum logic in this system [11]. Optical lattices [12], created by the AC Stark shift in laser standing waves tuned many thousand linewidths from atomic transition, are well suited for studies of quantum coherent dynamics, in part due to low rates of decoherence and in part due to the flexibility with which the optical potential can be designed [13]. If the intrinsic rate of decoherence is sufficiently suppressed, it is further possible to engineer dissipation back into the system in a well controlled manner through a combination of noise on the potentials, photon scattering and Raman sideband cooling [14]. Last but not least, one can hope to apply a range of quantum control techniques, including pure state preparation, controlled unitary evolution and quantum state reconstruction. Earlier work on quantum transport in optical lattices has explored a number of related phenomena such as Bloch oscillations and Wannier-Stark ladders [15], Landau-Zener tunneling [16], and tunneling in optical gauge potentials [17]. The coherent dynamics studied in those experiments, however, involved shallow potentials and a continuum of Bloch states, rather than the discrete two-level dynamics which can be achieved in a deep double-well potential.

Our atom/optical lattice system has been discussed in detail in [13], and only the most important features are summarized here. The lattice is formed by two counterpropagating laser beams with linear polarizations at a relative angle $\theta$ (1D lin-$\theta$-lin configuration), detuned $\sim 3000\Gamma$ below the Cesium $6S_{1/2}(F=4) \to 6P_{3/2}(F=5)$ transition. The resulting light field consists of a pair of $\sigma_+$ and $\sigma_-$ polarized standing waves with a relative spatial phase determined by $\theta$. When the laser detuning is much larger than the excited-state hyperfine splitting (as is the case here) the lattice potential can be written in terms of a scalar potential (proportional to the field intensity) and a fictitious magnetic field (proportional to the field ellipticity) interacting with the magnetic moment $\hat{\vec{\mu}} = -g_F \mu_B \hat{\mathbf{F}}$, where $\hat{\mathbf{F}}$ is the vector angular momentum operator for the hyperfine state $F$. If we include an external magnetic field $\mathbf{B}$ the resulting potential is

$$\hat{U}(z) = U_J(z) + g_F \mu_B \hat{\mathbf{F}} \cdot \mathbf{B}_{eff}(z),$$
$$U_J(z) = \frac{4U_1}{3}[1 + \cos\theta \cos(2k_L z)],$$
$$\mathbf{B}_{eff}(z) = -\frac{2U_1}{3\mu_B}\sin\theta \cos(2k_L z)\mathbf{e}_z + \mathbf{B},$$

where $U_1$ is the light shift produced by a single lattice beam driving a transition with unit oscillator strength.

One typically defines *diabatic* and *adiabatic* potentials as the diagonal elements (in the basis $\{|m_F\rangle\}$) and eigenvalues of $\hat{U}(z)$ respectively; the former govern the motion when the internal atomic state is independent of $z$ (e. g. in a lattice with no coupling between different $|m_F\rangle$), the latter when it adiabatically follows the direction of the local magnetic field (the Born-Oppenheimer approximation). For our parameters the lowest adiabatic potential (Fig. 1) forms a periodic array of double-wells. The lattice polarizations on the two sides of the well are predominantly $\sigma_+$ and $\sigma_-$ and the eigenstates of $\hat{U}(z)$ in these regions have predominantly $m_F > 0$ and $m_F < 0$ character, so that motion from one side of the well to the other is accompanied by rotation of the spin. The spin thus acts as a "meter" through which one can measure the evolution of the center-of-mass atomic wave packet.

For our system the Born-Oppenheimer approximation breaks down and one cannot describe the dynamics in terms of a particle moving on the adiabatic potential. We solve instead for the *exact* energy spectrum (band structure) and stationary states of the complete lattice Hamiltonian. For the parameters depicted in Fig. 1 the two lowest bands are split by a small energy $\hbar\Omega$, much less than the separation to the next excited bands. In addition, the negligible band curvature indicates that tunneling between different double-wells is unimportant. It is then possible to restrict the system dynamics to the subspace spanned by the Wannier spinors $|\psi_S\rangle$ and $|\psi_A\rangle$, corresponding to the symmetric/antisymmetric ground doublet of individual double-wells. We can recast the problem in familiar terms by defining left and right localized spinor wavepackets $|\psi_L\rangle = (|\psi_S\rangle + |\psi_A\rangle)/\sqrt{2}$ and $|\psi_R\rangle = (|\psi_S\rangle - |\psi_A\rangle)/\sqrt{2}$, and see immediately that the system will Rabi oscillate between these at frequency $\Omega$ if initially prepared in e. g. $|\psi_L\rangle$.

Fig. 2a shows the spatial probability distribution obtained by tracing over the internal state, while Fig. 2b shows the magnetic populations obtained by tracing over the center-of-mass coordinate, at different times during this oscillation. From Fig. 2a it is evident that $|\psi_L\rangle$ and $|\psi_R\rangle$ can be nearly distinguished by their mesoscopic center-of-mass distributions.

We prepare atoms in the double-well optical lattice and follow their quantum coherent dynamics using established techniques of laser cooling and trapping. First, a standard MOT/3D molasses is used to prepare a sample of $\sim 10^6$ Cesium atoms with a temperature of $\sim 4$ μK and a $\sim 200$ μm RMS radius. The atoms are cooled further in a near-resonance 1D lin-θ-lin lattice and then adiabatically transferred to the far-off-resonance 1D lin-θ-lin double-well lattice. The two 1D lattices are oriented vertically, which allows us to measure the atomic momentum distributions by time-of-flight analysis, and the magnetic populations by Stern-Gerlach analysis [18]. Great care is taken to assure that the lattice polarizations are linear and at the appropriate angle, and that the background magnetic field is reduced to $\sim 0.3$ mG or better. Once in the far-off-resonance lattice the atoms are optically pumped to the $|m_F = 4\rangle$ magnetic sublevel. We then select the motional ground state in the $m_F = 4$ potential by lowering the lattice depth to a point where only the lowest band is bound, and accelerating the lattice at $300$ m/s$^2$ for 1.5 ms to allow atoms in higher bands to escape [16]. The motional ground state selection is done in the presence of a large external $B_z$ to lift degeneracies between different potentials and prevent precession of the magnetic moment. When the population in higher bands has been eliminated we increase the lattice depth to the value used in the experiment and change the acceleration so the lattice reference frame is in free-fall. This prepares roughly 90% of the atomic population in the lowest band of the $m_F = 4$ potential, estimated from the measured momentum spread and magnetic populations. To initiate a clean Rabi oscillation we adiabatically connect this state to a localized state in a symmetric double-well potential, by ramping $B_x$ from zero to its desired value in $250$ μs, then ramping $B_z$ to zero in $70$ μs. To create the desired localized state the final turn-off of $B_z$ must be fast compared to the ground-doublet splitting, but slow compared to the separation between the ground doublet and higher bands. We have checked, by numerical integration of the time-dependent Schrödinger equation using the full lattice Hamiltonian, that this requirement can be met over a wide range of parameters including those used here.

Rabi oscillation between $|\psi_L\rangle$ and $|\psi_R\rangle$ are detected by measuring the magnetic populations. Fig. 3 shows a typical oscillation of the average magnetization as a function of time. Our data fits very well to an exponentially damped sinusoid, and we can extract good measures for the Rabi frequency over a wide range of parameters. Figs. 4a and 4b show the measured frequencies versus the single beam light shift $U_1$ and transverse magnetic field $B_x$, together with the ground doublet splitting predicted by band structure calculations. To carry out a direct theory/experiment comparison we independently measure $U_1$ to within ±2% [19] from parametric wavepacket oscillations, and the external $B_x$ to within ±1% from Larmor precession of the magnetic moment. Excellent agreement is observed, especially if we allow for a ~4% systematic underestimate of $U_1$. Fig. 4c shows the variation of the Rabi frequency versus $B_z$, which changes the energy asymmetry (detuning) of the two-level system. The observed dependence is characteristic of two-level dynamics and confirms that our system is effectively restricted to the ground doublet.

We have also the possibility to examine the magnetic populations in detail at different times during the Rabi oscillation. Fig. 2c shows typical values at $t \sim 0$, $t = \tau/4$ and $t = \tau/2$, where $\tau = 2\pi/\Omega$ is the Rabi period. Generally, we find qualitative agreement with the results of a band structure calculation, though the experiment shows a somewhat smaller net magnetization than expected. This might indicate that the initial state is in fact an incoherent superposition with ~80% population in $|\psi_L\rangle$ and ~20% population in $|\psi_R\rangle$. There is also a slight modification of the measured populations due to a small degree of adiabatic following as we turn on a magnetic field to define the quantization axis for our Stern-Gerlach measurement. This effect is responsible for the deviation from the expected mirror symmetry of $|\psi_L\rangle$ and $|\psi_R\rangle$ around $m_F = 0$. Of particular interest is the spinor wavepacket at $t = \tau/4$, where $|\psi(t)\rangle \propto |\psi_L\rangle - i|\psi_R\rangle$. Within the limits just mentioned we measure magnetic populations consistent with this delocalized superposition state, though the populations by themselves do not allow us to distinguish between a coherent superposition and an incoherent mixture of $|\psi_L\rangle$ and $|\psi_R\rangle$. Evidence of the coherence comes instead from the persistence of Rabi oscillations at later times.

As illustrated by the data in Fig. 3, we find that the amplitude of Rabi oscillations in the double well decays with a time constant of a few hundred microseconds. We estimate the timescale for decoherence from photon scattering to be ~1 ms, which is too slow to account for the observed damping. We have looked at Rabi oscillations also in a lattice tuned above atomic resonance, where the coherent dynamics is identical but the rate of photon scattering reduced by a factor of two to three. In practice we see no significant difference in the decay rate. This suggests that the amplitude decay is caused by dephasing of the Rabi oscillations, which occurs because different atoms in the ensemble see a slightly different lattice environment. The most likely cause is an estimated ~5% variation of the lattice beam intensities, which is reasonably consistent with the observed dephasing time. The rapid dephasing underscores the fragile nature of highly entangled states of atomic internal and external degrees of freedom, and suggests that proposals for quantum information processing [11] should seek clean separation of spin and center-of-mass motion.

In summary we have observed Rabi oscillations of atomic spinor wavepackets between distinguishable mesoscopic states in the optical double-well potentials of a far-off-resonance 1D lin-θ-lin optical lattice. We have taken extensive data for a relative polarization angle $\theta = 80°$, plus additional data at $\theta = 85°$ (not shown here). Both data sets show Rabi frequencies in excellent agreement with the predictions of band structure calculations. The persistence of oscillation for at least a few Rabi periods indicates that quantum coherent superpositions of the left- and right- localized states occur at certain times. Furthermore, the decay of these oscillations is most likely not caused by intrinsic decoherence from the scattering of lattice photons, but rather by inhomogeneities in the lattice potential across the atomic sample. In principle such dephasing can be reversed by spin-echo techniques similar to those employed in nuclear magnetic resonance, and we are in the process of setting up an improved experiment where this possibility can be explored. With improved lattice homogeneity, larger lattice detuning and echo techniques we hope to explore coherent dynamics on timescales much longer than the Rabi period. One might then reintroduce dissipation in a controlled fashion and study the fundamental process of decoherence, as well as the transition from quantum coherent to classical dynamics. This last aspect is especially intriguing, as the coupled spin-motion Hamiltonian associated with this system can be mapped onto the Tavis-Cummings model without the rotating wave approximation, whose classical counterpart is known to exhibit deterministic chaos [20]. The flexibility associated with our system should allow us to study the effects of decoherence on the emergent nonlinear behavior [21]. Additional phenomena can be studied with the addition of coherent driving fields, such as the coherent suppression of tunneling [22].

We thank Paul Alsing, Kit-Iu Cheong, Shohini Ghose and Gerd Klose for many helpful discussions. PSJ was supported by NSF, ARO and JSOP. IHD was supported by NSF.

Fig. 1. Lowest two adiabatic potentials (thick curves) and six energy bands (thin lines) for a 1D lin- -lin optical lattice, for parameters $U_1 = 84 E_R$, $= 80°$, $B_x = 85$ mC and $B_z = 0$ mG.

Fig. 2. Spinor wavepackets in the optical double-well during a Rabi oscillation. Lattice parameters are identical to Fig. 1. (a) Center-of-mass probability density calculated from the Wannier states. The dotted curve in the plot for $t = /2$ indicates the distribution at $t = 0$, and shows the minimal spatial overlap of the left and right localized wavepackets. (b) Magnetic populations calculated from the Wannier states. (c) Experimentally measured magnetic populations. The preparation of the initial state is not instantaneous on the timescale of the Rabi oscillation and we cannot easily assign an effective $t = 0$ for the experiment. The first row therefore shows calculated and measured distributions at a slightly later time, $t = /10$.

Fig. 3. Typical magnetization oscillation as a function of time, for parameters identical to Fig. 1. The solid line is a fit to a exponentially decaying sinusoid.

Fig. 4. Measured Rabi frequencies for the double well system. Base lattice parameters are identical to Fig. 1, except for one parameter as indicated in the plots. (a) versus single beam light shift $U_1$, (b) versus $B_x$ and (c) versus $B_z$. Open (filled) circles indicate data taken for a lattice tuned below (above) resonance. The solid curves show the ground doublet splitting from bandstructure calculations, with no free parameters, the dashed curve shows the same splitting with a 4% increase in $U_1$ relative to our best independent estimate.

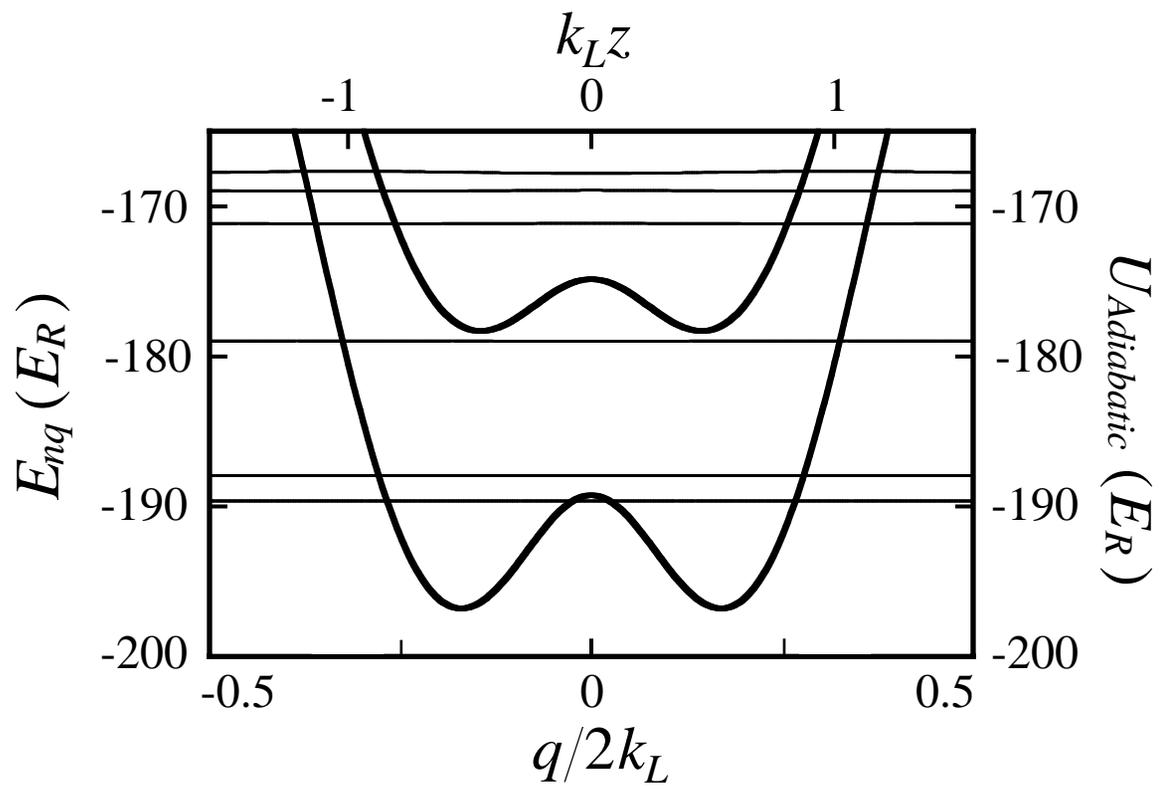

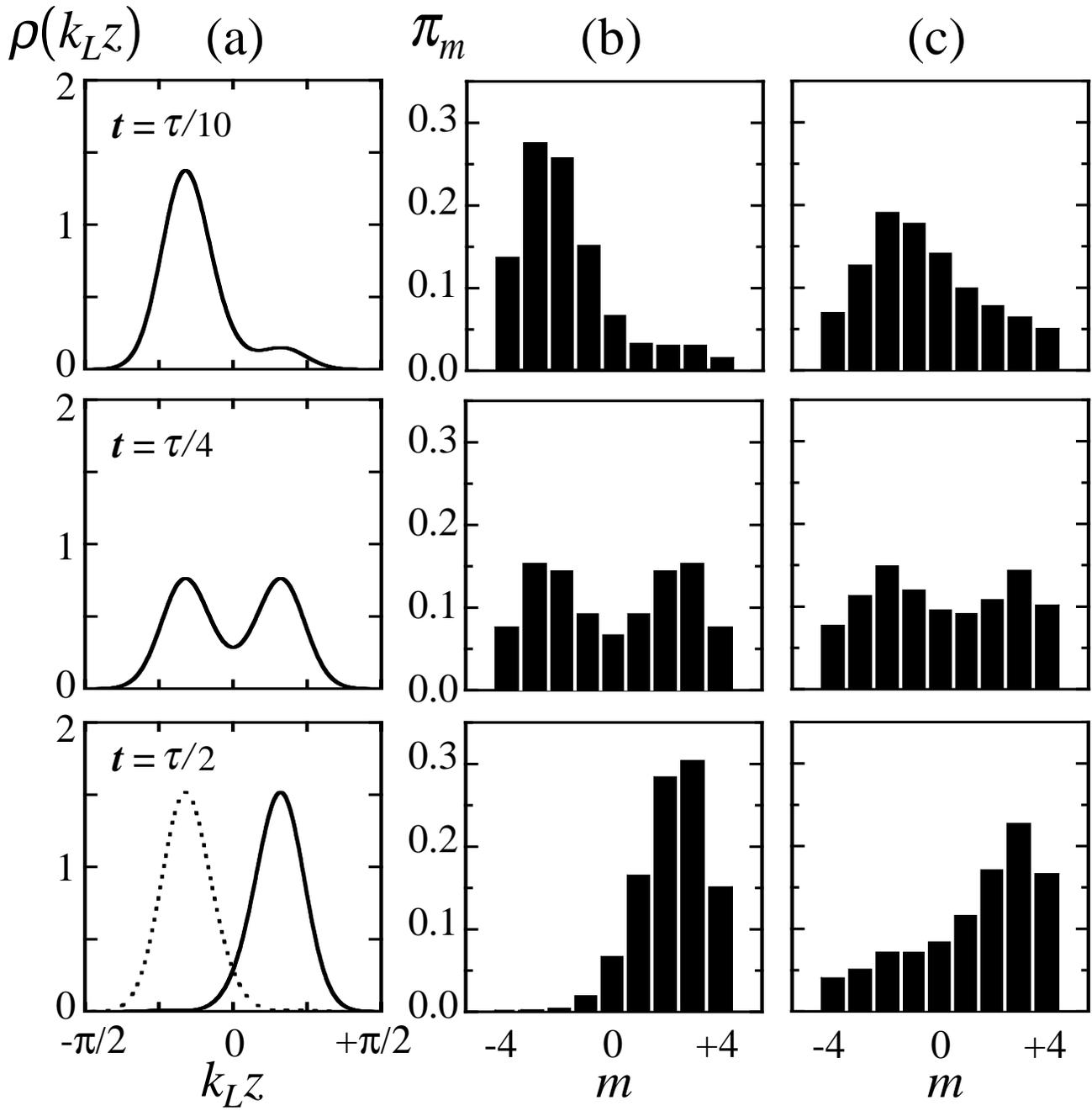

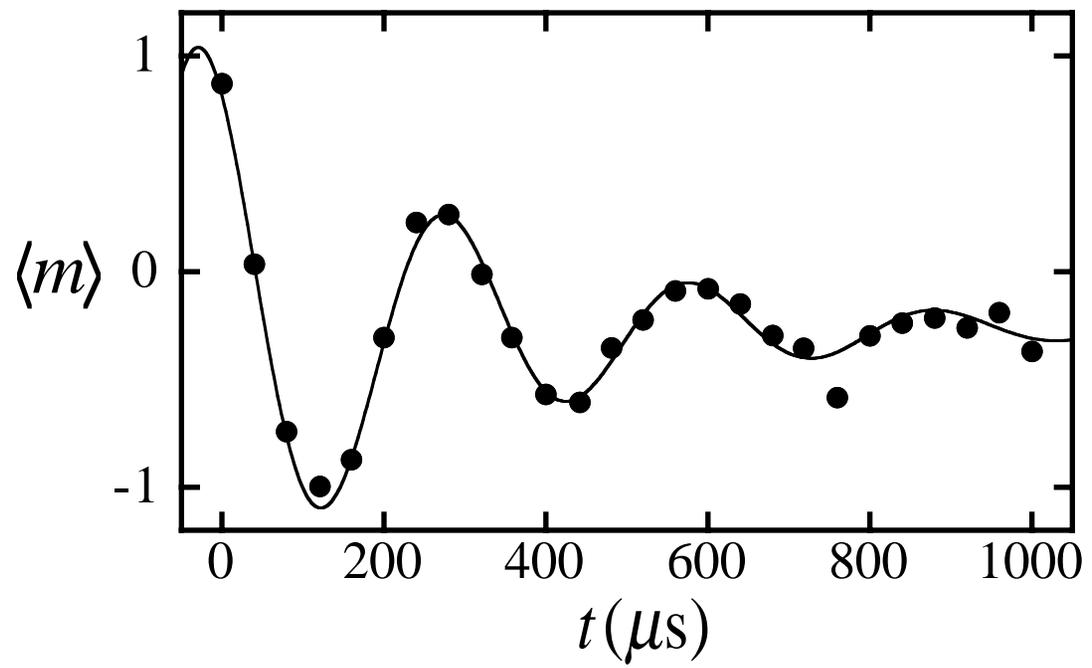

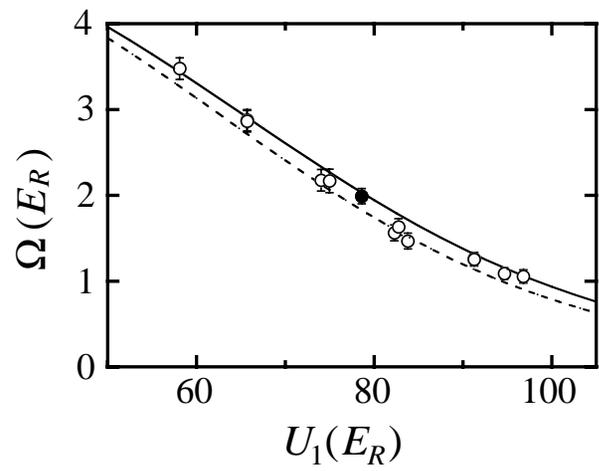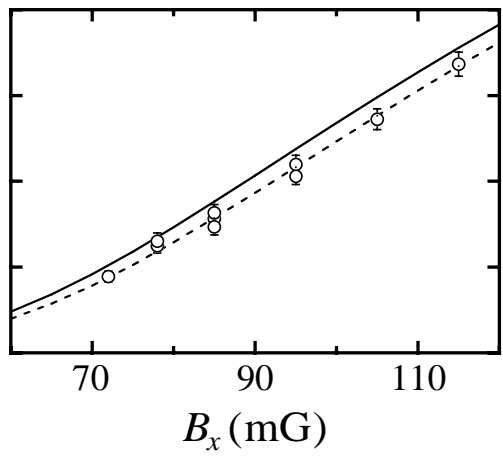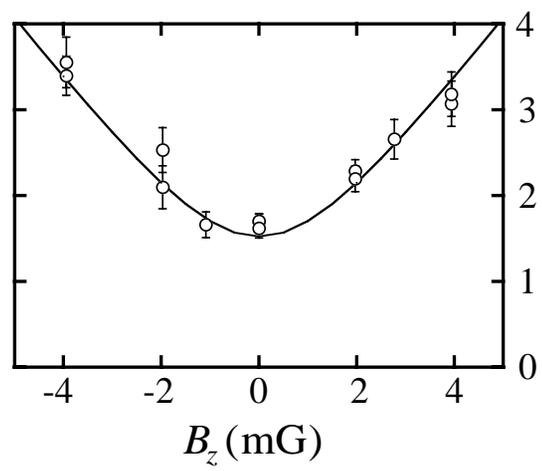